\documentclass[conference]{IEEEtran}
\IEEEoverridecommandlockouts
\usepackage{cite}
\usepackage{amsmath,amssymb,amsfonts}
\usepackage{algorithmic}
\usepackage{graphicx}
\usepackage{textcomp}
\usepackage{xcolor}
\usepackage{siunitx}
\usepackage{multirow}
\usepackage{subcaption}
\def\BibTeX{{\rm B\kern-.05em{\sc i\kern-.025em b}\kern-.08em
    T\kern-.1667em\lower.7ex\hbox{E}\kern-.125emX}}
\begin{document}

\title{Fine-Grained Coordinated OFDMA With Fiber Backhaul Enabled by openwifi and White Rabbit
\thanks{This work is partially funded by the Flemish FWO SBO
S006224N SMART-STRIP, Methusalem SHAPE: Next Generation Wireless Networks under grant 20-MET-004, and
HORIZON-ERC-POC DCAP (Distributed Coordinated Access Point) with project number 101158037.}
}

\author{\IEEEauthorblockN{Thijs Havinga, Xianjun Jiao, Wei Liu, Baiheng Chen, Robbe Gaeremynck and Ingrid Moerman}
\IEEEauthorblockA{IDLab, \textit{Department of Information Technology}, Ghent University - imec, Ghent, Belgium \\
Email: \{Thijs.Havinga, Xianjun.Jiao, Wei.Liu, Colvin.Chen, Robbe.Gaeremynck, Ingrid.Moerman\}@UGent.be
}}

\maketitle

\begin{abstract}
Proper coordination is needed to guarantee the performance of wireless networks in dense deployments. Contention-based systems suffer badly in terms of latency when multiple devices compete for the same resources. Coordinated Orthogonal Frequency Division Multiple Access (Co-OFDMA) is proposed for Wi-Fi 8 to remedy this, as it enables multiple Access Points (APs) to share spectrum more efficiently.
However, fine-grained resource allocation\textemdash namely within 20MHz bandwidth\textemdash is argued to be impractical due to the over-the-air scheduling overhead and complexity in terms of physical layer signaling. A wired backhaul mitigates the need for over-the-air scheduling and synchronization, and it allows for coordination even if APs are not in each others' range. Furthermore, it forms the basis for more advanced multi-AP coordination schemes like coordinated beamforming and joint transmission. In this work we demonstrate the realization of Wi-Fi 6 compliant fine-grained Co-OFDMA using a fiber backhaul, enabled by the open-source platforms openwifi and White Rabbit. 
We show that the performance in terms of carrier frequency offset pre-compensation and time synchronization between two APs exceeds related wireless standard requirements. Furthermore, the quality of the received constellation of the Co-OFDMA frame as reported by a wireless connectivity tester is better than individual frames sent by the APs.
\end{abstract}

\begin{IEEEkeywords}
OFDMA, Wi-Fi, Multi-AP Coordination, latency, reliability
\end{IEEEkeywords}

\section{Introduction}
In addition to throughput, latency and reliability have become crucial metrics for wireless communications. Applications such as industrial automation, robotic control, audio streaming, and virtual reality are particularly sensitive to high latency and jitter. It is therefore that the task group working on the IEEE 802.11bn standard, i.e. Wi-Fi 8, focuses on Ultra High Reliability (UHR) \cite{Galati-Giordano24}. Since Wi-Fi is working in unlicensed spectrum and uses contention-based access, this is a challenging task. For example, there are only three non-overlapping \SI{20}{MHz} channels available in the \SI{2.4}{GHz} band, which is still in active use due to its favourable propagation characteristics and compatibility with legacy devices.


One of the primary features for UHR is Multi-AP Coordination (MAPC), in which APs with overlapping coverage areas coordinate to achieve better performance in dense deployments such as automated factories. There are several flavors of MAPC, such as Coordinated Spatial Reuse (Co-SR), Beamforming (Co-BF), Time Division Multiple Access (Co-TDMA), Orthogonal Frequency Division Multiple Access OFDMA (Co-OFDMA) and Joint Transmission (JT). Using Co-SR, APs coordinate their transmit power to allow for simultaneous transmissions while limiting interference of stations (STAs) close the the AP. Simultaneous transmissions are also achieved using Co-BF, where APs send spatial nulls to the STAs they do not need to serve. With Co-TDMA, multiple APs share time resources during a transmit opportunity, while using JT, the signals of multiple APs constructively overlap at the receiver to increase its signal quality.

In this work, we focus on downlink Co-OFDMA. OFDMA was introduced to Wi-Fi 6 in order to serve multiple users at the same time, each using a subset of the available subcarriers in the channel. Each subset is called a Resource Unit (RU) and the allocation of RUs is the task of the AP for both downlink and uplink traffic. With Co-OFDMA, multiple APs schedule the RUs jointly in order to reduce contention and occupy the spectrum more efficiently, and is therefore a promising technique to meet latency constraints as presented in \cite{Imputatio24}. The concept is similar to Fractional Frequency Reuse (FFR), which is a Coordinated Multi-Point (CoMP) technique used in cellular networks \cite{Lei07}. Herein, users at the cell edge will be served with a frequency portion that does not overlap with another cell, whereas for users in the cell center, the full frequency range can be reused.

A downside of Co-OFDMA, and MAPC in general, is the large overhead required to efficiently schedule the transmissions. APs need to exchange information about their traffic buffer status, and assess the channel quality to determine whether Co-OFDMA should be used. The authors of \cite{Lacalle22} consider Co-OFDMA as an important enabler for reliable industrial Wi-Fi. They propose a method to limit the control overhead by performing virtual sounding, i.e. instead of obtaining the Channel State Information (CSI) between every link in each cycle, only some links are measured and others are interpolated using old measurements. In a proposal for the IEEE 802.11bn standard \cite{IEEE23_0768}, it is argued that fine-grained Co-OFDMA using small RUs, namely less than \SI{20}{MHz}, is not feasible in Wi-Fi due to the scheduling overhead and the need for a preamble redesign. However, the work in \cite{Lacalle22} shows that Co-OFDMA is even effective with RU bandwidths of \SI{2}{MHz}.

In this work, we first motivate why fine-grained Co-OFDMA is important for UHR, and justify the use of a wired backhaul to achieve this. Then we discuss the system requirements, after which we describe our methodology to realize fine-grained Co-OFDMA compatible with the Wi-Fi 6 standard. Next, we present a performance evaluation using an experimental set-up, after which we discuss this paper and future work.

\section{Motivation}
OFDMA in Wi-Fi 6 allows for up to nine users per \SI{20}{MHz} bandwidth, with three different partial-bandwidth RU widths (106, 52 or 26 subcarriers). From Wi-Fi 7 onwards, multiple RUs can be combined to serve one user to enhance the flexibility, e.g., it allows for combining multiple RUs for a high-rate user, together with a small RU reserved for a low-rate, low-latency user.

Restricting the minimum RU width for Co-OFDMA to \SI{20}{MHz} would greatly reduce its applicability for lower-rate, low-latency and reliable wireless systems. These systems typically do not benefit from wider bandwidths, as smaller payloads require aggregation to reduce contention and preamble time overhead. However, aggregation inherently increases latency, making it unsuitable for applications demanding low-latency performance. Furthermore, the increased noise and interference associated with wider bandwidths due to transmit power limitations reduce the reliability of the system. Moreover, narrowband cross-technology interference such as IEEE 802.15.4 or Bluetooth can also be mitigated using OFDMA as we have shown in our previous work \cite{Havinga24}.

\begin{figure}
    \centering
    \includegraphics[width=0.8\linewidth]{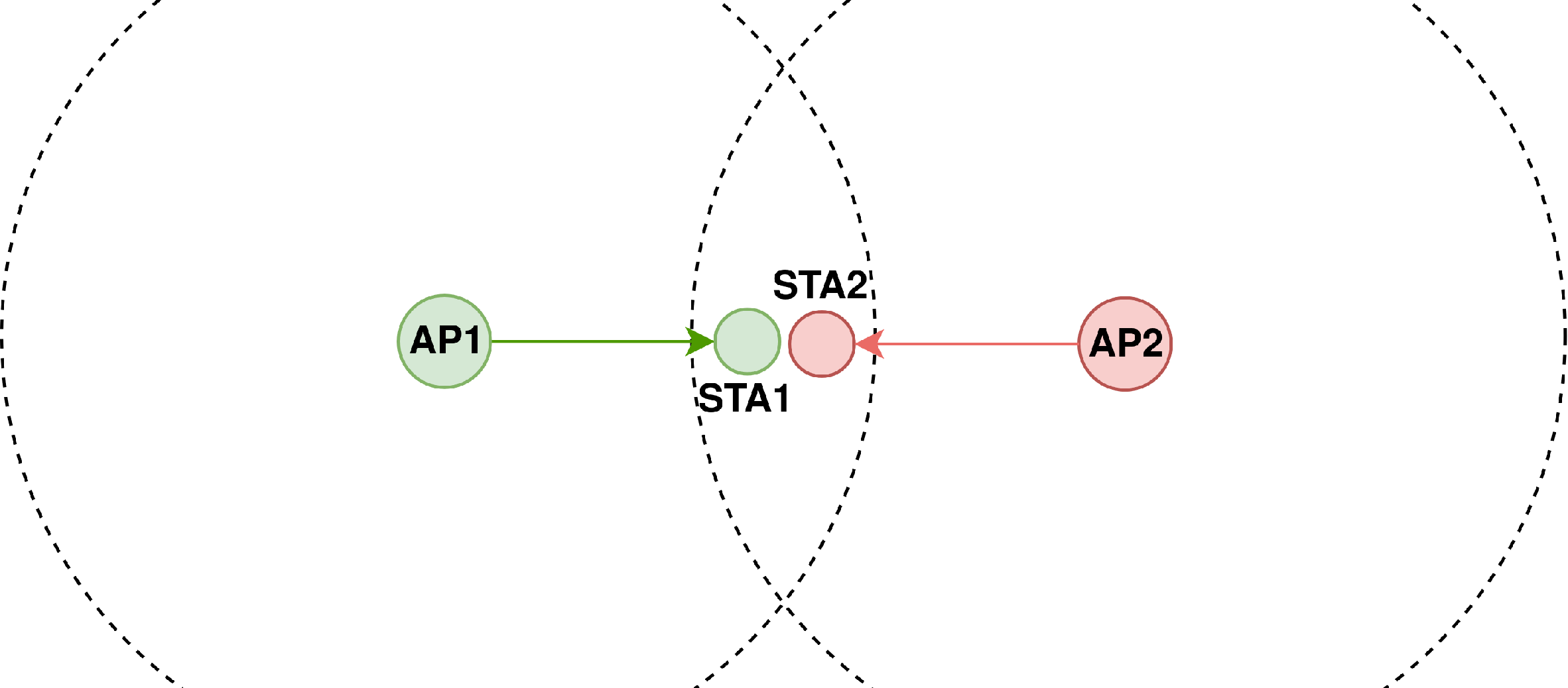}
    \caption{The scenario where AP1 and AP2 cannot hear each other, but simultaneous downlink packets to STA1 and STA2 will collide if the channels overlap.}
    \label{fig:C_OFDMA_scenario}
\end{figure}

We propose a wired backhaul to solve the issue of excessive scheduling overhead when considering sub-\SI{20}{MHz} Co-OFDMA, while still being a viable solution for cost-effective deployments. Furthermore, it brings an additional benefit of Co-OFDMA in a scenario where APs are hidden nodes to each other as shown in Figure \ref{fig:C_OFDMA_scenario}. 
In this scenario the coordination between APs cannot be done over-the-air. The APs might use different Service Set Identifiers (SSIDs), but even for the same SSID the coverage area of APs often overlaps for better connectivity. Choosing non-overlapping channels is not always possible due to limited available spectrum, or it is not desired due to the increase in latency during inter-channel roaming. Since packet errors are costly and lead to high latency and jitter due to retransmissions and increased contention windows, protection methods are needed. In the existing Wi-Fi standard, this is achieved by letting the AP send a Request-to-Send (RTS) frame, upon which the STA has to respond with a Clear-to-Send (CTS) frame to notify the other AP it is about to transmit.
However, there is a trade-off between efficiency gain due to reduced frame loss and efficiency loss due to extra signaling overhead, the latter being dominant for small payloads. 

\begin{figure}
    \centering
    \includegraphics[width=\linewidth]{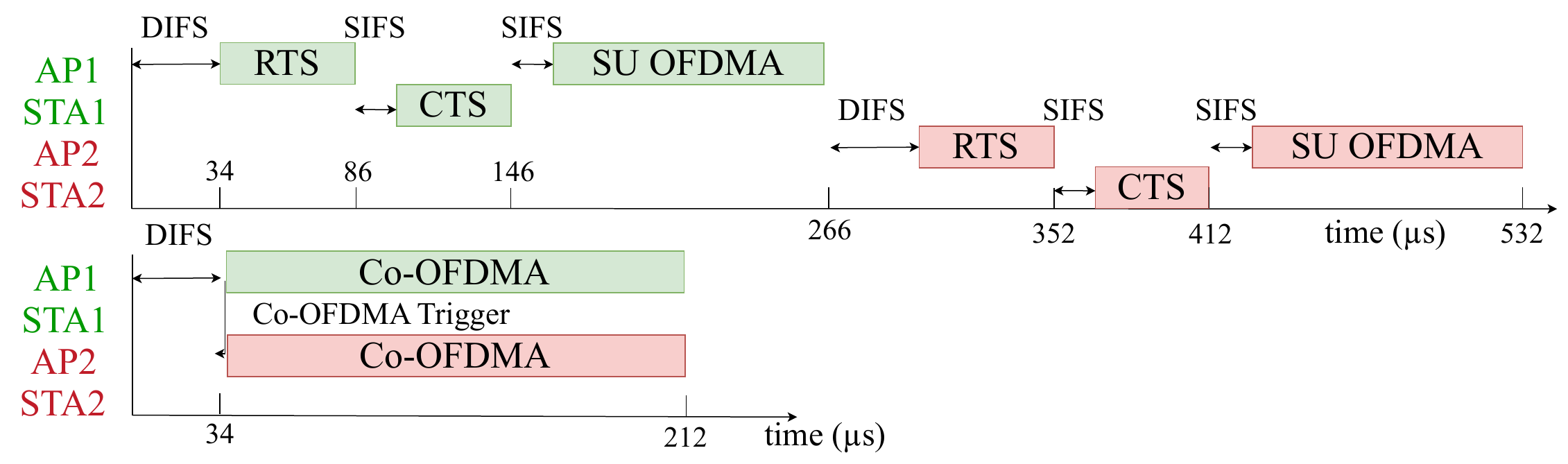}
    \caption{Time schedule of traditional RTS/CTS-based method (top) versus Co-OFDMA (bottom) for  downlink packets from AP1 and AP2 (Figure \ref{fig:C_OFDMA_scenario}).}
    \label{fig:Co_OFDMA_schedule}
\end{figure}

For example, for the scenario of Figure \ref{fig:C_OFDMA_scenario} where both APs have a 500-byte downlink packet to send to their associated STA on the same channel, we illustrate the time schedule as shown in the top axis of Figure \ref{fig:Co_OFDMA_schedule}. When using Modulation and Coding Scheme (MCS) 7, only 4 OFDM symbols are needed for a Single User (SU) packet when a bandwidth of \SI{20}{MHz} is used. This leads to an airtime of \SI{104}{\us}, including the preambles. Furthermore, the duration of an RTS packet is \SI{52}{\us} and the CTS is \SI{44}{\us} long. Before transmitting a downlink packet, an AP has to contend for the medium to send an RTS packet. The minimum contention period is equal to the time of the DCF (Distributed Coordination Function) Interframe Space (DIFS), which is \SI{34}{\us} as shown in Figure \ref{fig:Co_OFDMA_schedule}. On top of this, more backoff time could be added depending on the contention window size. Upon succesfull reception of the RTS, the STA has to respond with a CTS after the Short Interframe Space (SIFS), which is \SI{16}{\us}. Afterwards, the AP can send the SU packet following another SIFS. In total, in the best case (no backoff needed) \SI{532}{\us} is needed for the two downlink packets to be transmitted.
If Co-OFDMA is used in this scenario, the APs can use a Multi-User (MU) packet with two 106-tone RUs where each AP transmits over a single RU. This leads to 7 OFDM symbols per user, resulting in an airtime of \SI{176}{\us} including preamble. First, AP1 will contend for the medium as usual, and then send a Co-OFDMA trigger over the fiber link (denoted with \SI{2}{\us} latency). Both APs then start the Co-OFDMA packet at the same time, and the concurrent data transmission to two STAs finishes after only \SI{212}{\us}. 

Another benefit of Co-OFDMA is that the power spectral density and hence Signal-to-Noise Ratio (SNR) at the receiver can be increased. 
Namely, while the maximum transmit power per AP stays the same, APs only allocate power to the subcarriers they are assigned to, boosting power per subcarrier. This is a similar gain as achieved with uplink OFDMA. When taking the boosted received power into account, which is roughly \SI{3}{dB} for two APs, a higher MCS may be possible, which will lead to even shorter airtime and hence higher throughput for the Co-OFDMA packet.

\section{System Requirements}
The requirements of downlink fine-grained Co-OFDMA will be similar to regular uplink OFDMA, as with the latter, STAs transmit simultaneously to the AP using different RUs. Firstly, it is important to have a low-latency and low-jitter trigger, as the transmission should begin promptly after completing contention, and the transmissions from different APs should start coherently. Generally, as long as two signals with relative delay to each other are received within the Guard Interval (GI), it is ensured that no Inter-Symbol Interference (ISI) occurs, as the insertion of a Cyclic Prefix (CP) assures the orthogonality of subcarriers remains. Since the shortest GI is \SI{800}{ns}, for uplink OFDMA the transmission should start within $\pm$\SI{400}{ns} after the SIFS following the trigger frame as given by the IEEE 802.11ax standard \cite{80211ax}. We adopt the same time constraint for downlink OFDMA by different APs. 

Moreover, since the preamble, which is also used for carrier frequency offset (CFO) estimation, occupies the full 20MHz bandwidth, the APs have to synchronize their carrier frequencies. If a receiver does not compensate for CFO, it disrupts the orthogonality of the subcarriers, leading to Inter-Carrier Interference (ICI). When two devices transmit a preamble with different CFOs relative to the receiver, the receiver cannot estimate\textemdash and hence compensate for\textemdash the CFO accurately. For uplink OFDMA, the 10\% exceedance point of residual CFO error between AP and STA is set to \SI{350}{Hz} \cite{80211ax}, which we will also use as requirement for downlink Co-OFDMA.

CFO pre-compensation needs to be done for other MAPC techniques like Co-BF and JT as well, and for the latter it even needs phase coherency. To eliminate the need to synchronize the Local Oscillator (LO) of each AP, a coaxial cable could be used to distribute a common clock. In our previous work \cite{Chen25}, we have shown JT with two APs using a common clock, while the trigger was transferred via a general purpose input/output line. However, distributing a high-quality clock signal is challenging and expensive due to, among others, signal degradation and interference. Other methods include using a GPS disciplined oscillator \cite{Bui22}, however, this is expensive and usually not reliable indoors. NXP's Network Listening Module \cite{NXP} uses cellular networks, but this is also an expensive option and relies on other existing infrastructure. The work in \cite{Mangione24} argues that the LOs can be calibrated using a reference clock upfront, but this does not account for temperature and voltage changes, or long-term aging effects of the oscillator.

In our work we use a fiber backhaul to pre-compensate for CFO in order to benefit from the existing link used for scheduling and control packets, and potentially in the future for data sharing in JT.




\section{Methodology}
In order to realize fine-grained Co-OFDMA with a fiber backhaul, we need to modify the baseband processing of a Wi-Fi chip and implement the CFO pre-compensation and time synchronization over the fiber link. We realize this by combining two open-source projects --- openwifi and White Rabbit (WR).

Openwifi is an IEEE 802.11 transceiver running on software-defined radio \cite{jiao2020openwifi}, compatible with Linux' mac80211 subsystem. The baseband processing and low-MAC layer is implemented on an FPGA (also called Programmable Logic (PL)), while the higher MAC runs on the on-chip ARM processor (Processing System (PS)). The AD9361 RF front end interfaces with the PL and PS to send and receive the IQ samples.
In our previous work \cite{Aslam24}, we extended openwifi with OFDMA support according to the IEEE 802.11ax standard.

The White Rabbit Project is a platform developed by CERN to create high-precision time, frequency and phase synchronization between different devices over a fiber link \cite{Lipinski11}. It consists of three components, namely Precision Time Protocol (PTP), Synchronous Ethernet (SyncE) and Digital Dual Mixer Time Difference (DDMTD). PTP governs the time synchronization, SyncE handles syntonization (i.e. frequency alignment), while DDMTD performs phase detection. In a WR network, one node acts as master and the others as slaves, where the latter adjust their clock to the former. Using this, frequency transfer with a jitter of \SI{2}{ps} over a single link with \SI{5}{km} fiber cable is achieved. It might therefore even be suited for JT, which requires phase coherency. 

We based our work on the recent development of ``Light Rabbit", which enables to run White Rabbit on a commercially available System-on-Chip, namely the AMD ZCU102, without any additional hardware \cite{LightRabbit}. We consider a set-up with two APs in this paper, which are directly connected to each other using a fiber cable inserted into the on-board small form-factor pluggable transceiver. More APs can be interconnected when using a dedicated WR switch. 

\begin{figure}
    \centering
    \includegraphics[width=\linewidth]{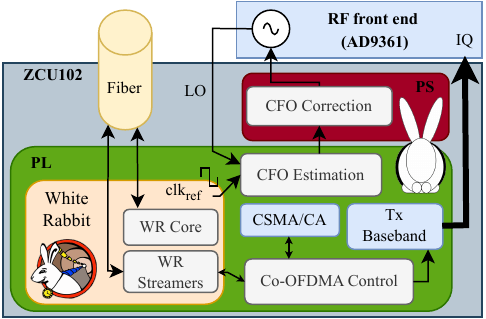}
    \caption{Schematic diagram of the integrated openwifi and White Rabbit design for Co-OFDMA on the AMD ZCU102.}
    \label{fig:openwifi_WhiteRabbit}
\end{figure}

A schematic diagram of the integration of openwifi and White Rabbit to support Co-OFDMA is shown in Figure \ref{fig:openwifi_WhiteRabbit}. It consists of several components that are discussed hereafter. Parts in light blue are existing modules of openwifi that are modified in this work, and the other parts are new modules.

\subsection{Carrier Frequency Offset Compensation}
The \SI{40}{MHz} crystal used by the AD9361 RF front end has a frequency stability of $\pm$\SI{18.1}{ppm} \cite{ad9361}. When applying this to a carrier frequency of \SI{2412}{MHz} (Wi-Fi channel 1), it can result in a CFO of $\pm$\SI{43.4}{kHz}. In order to compensate for this, the Digital-Controlled Crystal Oscillator (DCXO) can be tuned. Its resolution is \SI{0.0125}{ppm} in the worst case, meaning \SI{30.15}{Hz} at \SI{2412}{MHz}. Although it is possible to achieve a resolution of \SI{2.4}{Hz} by programming the fractional-N frequency synthesis of the LO as used by \cite{Mangione24}, this requires the phase-locked loop (PLL) to be re-locked, which could make a transmission or reception unstable. In contrast, the tuning of the DCXO can be performed without re-locking the PLL.

Once the White Rabbit link has been set up by the \textit{WR Core} inside the PL, the \SI{125}{MHz} syntonized reference clock ($\text{clk}_{\text{ref}}$) is used by the module \textit{CFO Estimation} inside the PL to estimate the clock frequency of the LO in the RF front-end. The module counts the number of clock cycles of the LO (at \SI{40}{MHz}) for a predetermined number of cycles of the reference clock (at \SI{125}{MHz}) and compares this with the expected number. 
To keep the measurement resolution below the tuning resolution of \SI{0.0125}{ppm}, the measurement period is set to 2.5 seconds (i.e. 312.5 million cycles at \SI{125}{MHz}), which leads to 100 million expected rising edges of the \SI{40}{MHz} clock, corresponding to \SI{0.01}{ppm} resolution. When applying this to \SI{2412}{MHz} carrier frequency, the maximum measurement error is \SI{24.12}{Hz}.
The difference between the expected and the actual count is written to a register that is shared with the PS, in which the \textit{CFO Correction} task runs periodically in the driver to tune the AD9361 DCXO via SPI commands. 

\subsection{Baseband adaptation in openwifi}
\begin{figure}
    \centering
    \includegraphics[width=\linewidth]{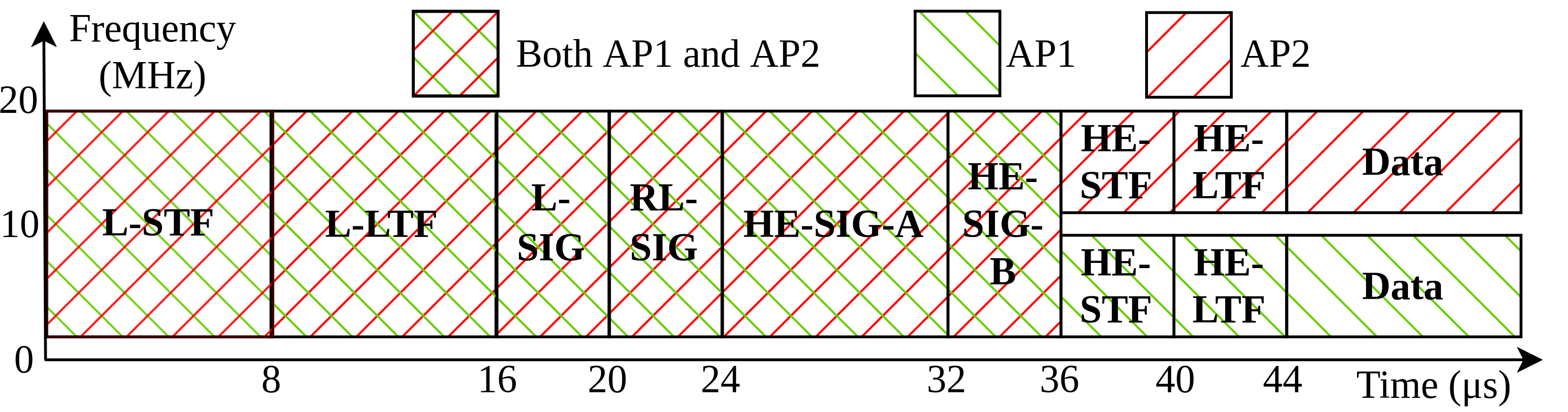}
    \caption{Fields of a 20MHz two-user HE-MU packet transmitted by two APs using fine-grained Co-OFDMA.}
    \label{fig:fields}
\end{figure}

Inside the PL, some changes are needed within the \textit{Tx Baseband} processing module. Firstly, to be compatible with the Wi-Fi 6 standard, the HE (High Efficiency) MU packet format is used. To ensure that the STAs can decode the Co-OFDMA frame, both APs have to compose the same full-bandwidth preamble extending to the HE-SIG-B field as shown in Figure \ref{fig:fields}.
This means that all the parameters in these fields should be known upfront at each AP. This includes, for example, the L-Length parameter representing the airtime of the packet in the L-SIG field, but also the Basic Service Set (BSS) color field in the HE-SIG-A. Normally, two APs will get a different BSS color and STAs will only decode packets corresponding to the color of their BSS. Therefore, APs participating in Co-OFDMA should use the same BSS color to maintain STA-side coherency. Similarly, since the STA-ID is only unique within a BSS, to support Co-OFDMA, 
the STAs must be assigned different STA-IDs in their respective BSS.
Furthermore, the OFDMA resource allocation information contained in the HE-SIG-B must therefore be shared across the fiber link. 

Next, the AP only occupies the RU that is assigned to it. This means that the power of the subcarriers of the channel estimation training fields and the data part that are used by another AP should be set to zero. Meanwhile, the power in the occupied RU needs to be scaled according to the ratio of occupied versus unoccupied subcarriers.

\subsection{Co-OFDMA Control}
For access to the medium, both APs follow their own Carrier Sense Multiple Access with Collision Avoidance (CSMA/CA) procedure. Once one AP gains access to the medium, it sends the trigger to the other AP via the fiber link and they both start transmitting immediately.
To avoid unintentional collisions by the other AP, which might sense its carrier busy, the receiving AP can decide to ignore the trigger and withdraw from the current Co-OFDMA frame. A similar mechanism is found for uplink OFDMA in the current Wi-Fi 6 standard, where the trigger frame contains the field ``CS required". If set, a STA will withdraw from the scheduled uplink OFDMA transmission if the medium is busy.

In our implementation, the \textit{CSMA/CA} module notifies the \textit{Co-OFDMA Control} module when it finishes its backoff. It then uses the \textit{WR Streamers} module to send the trigger over the fiber link. The WR Streamers provides an interface for low-latency transmission using a protocol on top of Ethernet. After the transmission time over the fiber link has passed, the \textit{Co-OFDMA Control} module notifies the \textit{Tx Baseband} module that it can start the processing to transmit the packet with the configuration as pre-negotiated or encoded in the trigger frame, and send the IQ samples to the RF front-end. 

\section{Performance Evaluation}
We perform the evaluation of fine-grained Co-OFDMA using two AMD ZCU102s which are interconnected by a fiber cable of around \SI{1}{m}. The evaluation is performed in terms of carrier frequency offset, time synchronization and error vector magnitude. 

\subsection{Carrier Frequency Offset}
\begin{figure}
    \centering
    \includegraphics[width=\linewidth]{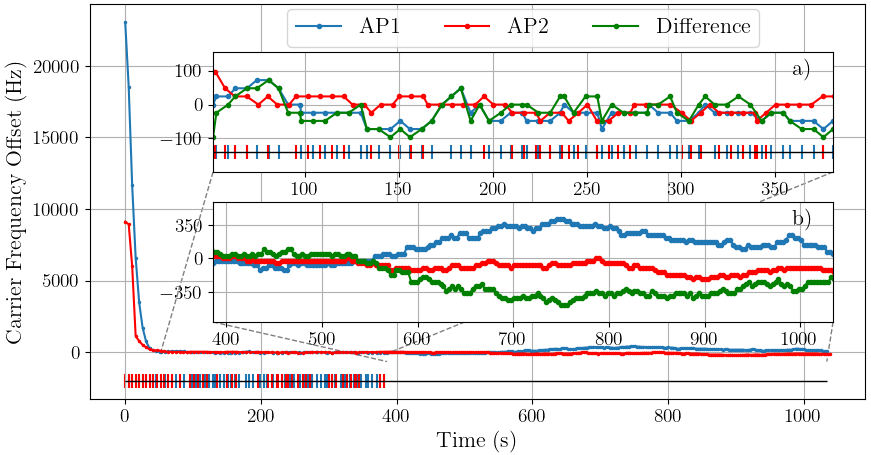}
    \caption{Measured CFO over time on two APs, with DCXO tuning events shown as vertical lines. The zoomed inset $a)$ is when tuning is enabled, and $b)$ when disabled.}
    \label{fig:cfo_time}
\end{figure}

Firstly we measure the CFO on each ZCU102 according to the WR reference clock after the WR link has been set up. Figure \ref{fig:cfo_time} shows the CFO over time, with the DCXO tuning events indicated by vertical lines. It shows that AP1 starts with an initial CFO of around \SI{23}{kHz}, while AP2 starts at around \SI{9}{kHz}. The inset $a)$ zooms in on the region with tuning after convergence, while the inset $b)$ zooms in to the area when tuning is disabled. From the global figure, we observe that after around \SI{50}{s}, both APs have tuned the oscillator. Inset $a)$ shows that the difference stays stable within $\pm$\SI{100}{Hz} range with a couple of tuning events per minute. Once the tuning is turned off, the LOs drift away from the WR reference clock. Their difference sometimes exceeds the limit of \SI{350}{Hz}, highlighting the need for continuous tuning.

\begin{figure}
    \centering
    \includegraphics[width=0.9\linewidth]{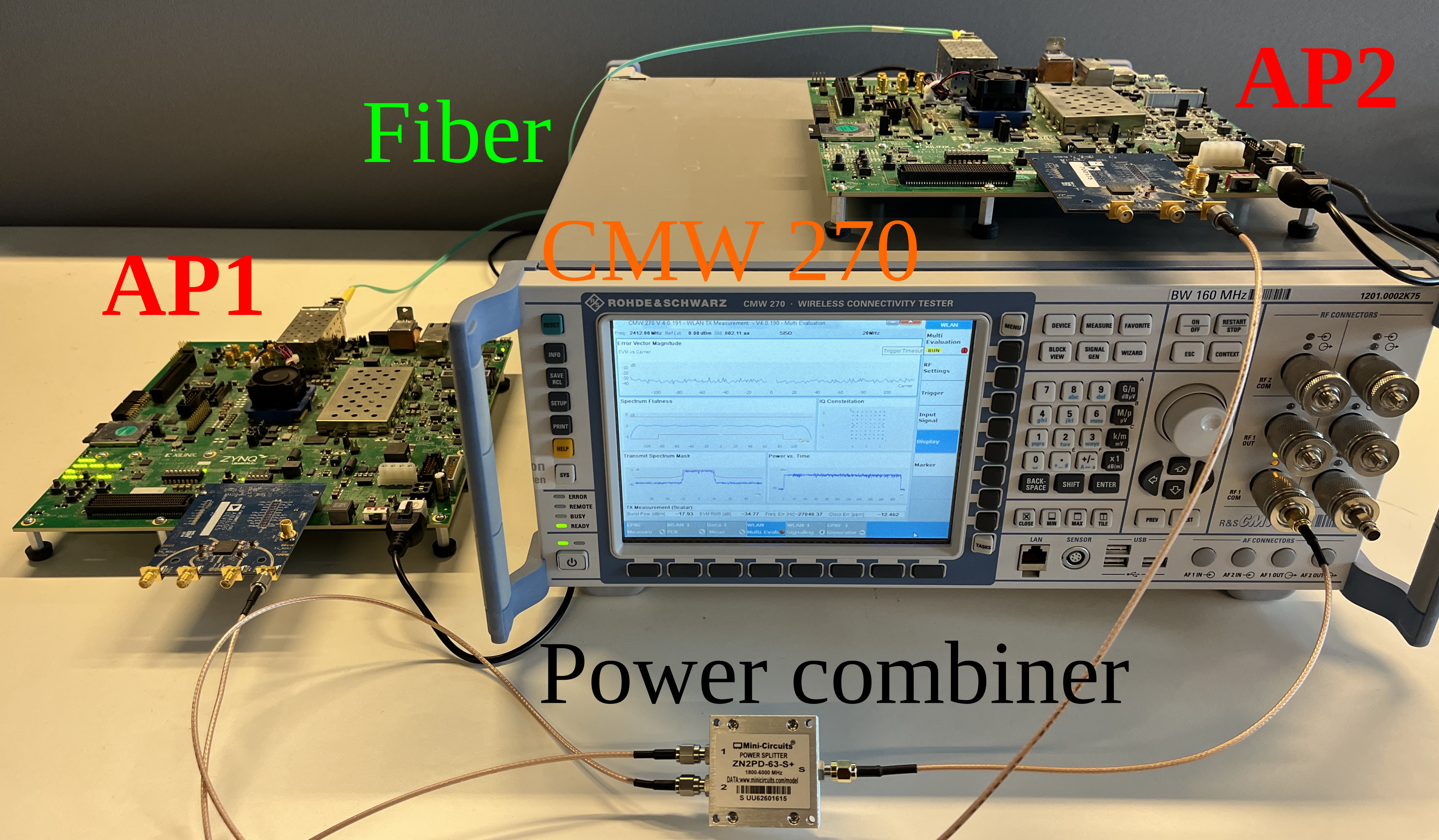}
    \caption{Measurement set-up of two APs interconnected with a fiber cable, their RF outputs are connected via a power combiner to the wireless tester.}
    \label{fig:meas_setup}
\end{figure}

Next, the output of the RF front ends of the ZCU102s are connected via a power combiner to the input of the R\&S\textsuperscript{\textregistered}CMW270 wireless connectivity tester as shown in Figure \ref{fig:meas_setup}.
We let AP1 and AP2 send fifty MU OFDMA packets independently quickly after each other by performing packet injection using a Linux user-space command. The CFO as determined by the tester is on average $-5587.87 \pm$ \SI{6.38}{Hz} for AP1 and $-5615.21 \pm$ \SI{6.20}{Hz} for AP2, meaning on average a remaining CFO of \SI{27.34}{Hz} between the two APs is achieved. This confirms the observation from Figure \ref{fig:cfo_time} and is more than sufficient according to the requirements of uplink OFDMA.

\subsection{Time synchronization}
\begin{figure*}[ht]
\centering
\begin{subfigure}{.5\linewidth}
  \centering
    \includegraphics[width=\linewidth]{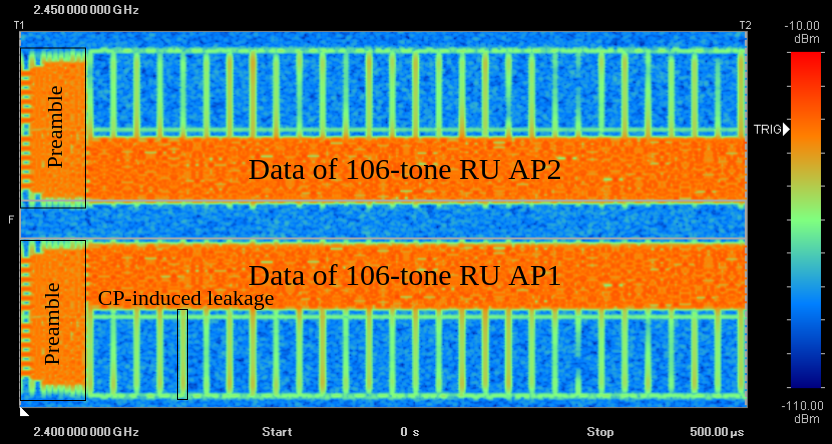}
    \caption{Span of 50MHz with AP1 set to channel 1 and AP2 to channel 6.}
    \label{fig:Co-OFDMA_two_chan}
\end{subfigure}%
\begin{subfigure}{.5\linewidth}
  \centering
    \includegraphics[width=\linewidth]{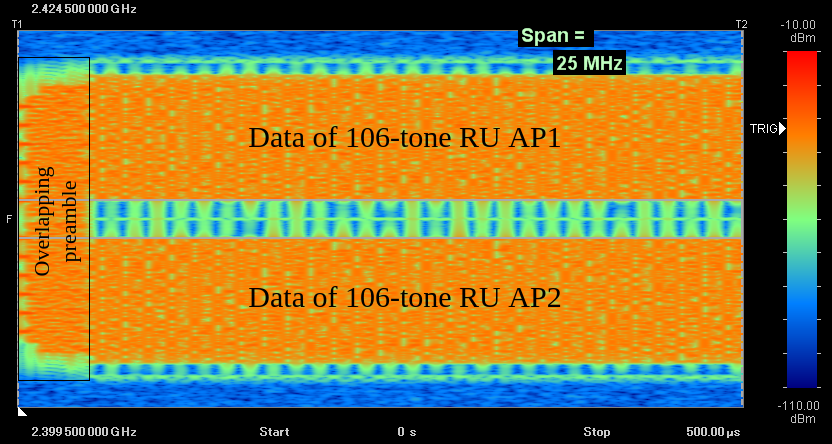}
    \caption{Span of 25MHz with AP1 and AP2 on channel 1.}
    \label{fig:Co-OFDMA_overlap}
\end{subfigure}
\caption{Spectograms with annotations showing a synchronized two-user fine-grained Co-OFDMA frame.}
\label{fig:test}
\end{figure*}
For time synchronization, we measure the variation in round-trip time of sending a trigger over the fiber link. In our previous work \cite{Chen25}, we have shown that the additional jitter caused by the baseband processing can be maintained within one clock cycle. The measurement is performed using 500 triggers started from AP1, and 500 triggers started form AP2. After the first AP sends the trigger, the other will send one back immediately after receiving it, such that the first can measure the time difference. The resulting round-trip time is on average \SI{3.834}{\us} with a standard deviation of \SI{14.9}{ns}. When setting the expected latency to an integer multiple of clock cycles of the WR system clock at \SI{62.5}{MHz}, the maximum deviation would be three clock cycles, corresponding to $\pm$\SI{48}{ns}. This is well below the required \SI{400}{ns} as defined by the IEEE 802.11ax standard for uplink OFDMA. 
The absolute one-way latency of around \SI{1.92}{\us} is not insignificant, however, it is lower than the receive-to-transmit turnaround time of \SI{5}{\us} as required by the IEEE 802.11 standard. Furthermore, the latency of the WR Streamers module is already identified as a source of possible improvement\footnote{https://ohwr.org/project/wr-cores/-/issues/93, accessed on April 1st, 2025.}, as the latency of a native fiber link could be within \SI{1}{\us}.

To visualize the spectrum of the Co-OFDMA transmission over time, the same set-up as shown in Figure \ref{fig:meas_setup} is used, but now the combined output is connected to an Anritsu MS2690A spectrum analyzer. First, AP1 was set to Wi-Fi channel 1 and AP2 to Wi-Fi channel 6 in order to capture the individual signals more clearly in Figure \ref{fig:Co-OFDMA_two_chan}. It can be seen that both APs first send an identical preamble at the same time, where the legacy part occupies the full \SI{20}{MHz} bandwidth. Afterwards, AP1 only transmits on the 106-tone RU 1, while AP2 only transmits on RU 2. Note that a transient artifact appears at the beginning of each OFDM symbol due to the discontinuity introduced by CP insertion, resulting in a time-domain signal that is not perfectly smooth. Next, in Figure \ref{fig:Co-OFDMA_overlap}, both APs are set to channel 1. The two full-bandwidth preambles now perfectly overlap and each AP sends the data on a distinct portion of the shared \SI{20}{MHz} channel.

\subsection{Error Vector Magnitude and Burst Power}
For the evaluation the Error Vector Magnitude (EVM) of the received constellation, we utilize the measurement set-up as shown in Figure \ref{fig:meas_setup}. First, 500 two-user MU OFDMA packets are sent by each AP individually, then another 500 packets are sent by both APs jointly using Co-OFDMA. The results from the tester are shown in Table \ref{tab:evm}. 
The average burst power of the Co-OFDMA packets is around \SI{3}{dB} higher than the individual packets, as is expected from the scaling each AP applies to its occupied RU. It can be seen that the EVM of Co-OFDMA is on average about \SI{1.4}{dB} lower than the packets sent by the individual APs, indicating improved receiving performance. It does not fully reach a \SI{3}{dB} improvement, suggesting that at high signal levels, transmitter noise becomes the dominant limiting factor, preventing a direct one-to-one improvement in received signal quality with increased power as argued in \cite{Chen25}.


\begin{table}[]
\centering
\captionsetup{width=0.9\columnwidth}
\caption{Error Vector Magnitude and Burst Power of the received MU and Co-OFDMA packets.}
\label{tab:evm}
\resizebox{0.9\columnwidth}{!}{%
\begin{tabular}{|l|ll|ll|}
\hline
\multirow{2}{*}{\textbf{}} & \multicolumn{2}{l|}{\textbf{EVM {[}dB{]}}} & \multicolumn{2}{l|}{\textbf{Burst Power {[}dBm{]}}} \\ \cline{2-5} 
                           & \multicolumn{1}{l|}{Average}        & Std. Dev.      & \multicolumn{1}{l|}{Average}            & Std. Dev.          \\ \hline
\textbf{MU AP1}                    & \multicolumn{1}{l|}{-34.77}     & 0.44     & \multicolumn{1}{l|}{-18.05}          & 0.01         \\ \hline
\textbf{MU AP2}                    & \multicolumn{1}{l|}{-35.22}     & 0.42     & \multicolumn{1}{l|}{-17.92}          & 0.01         \\ \hline
\textbf{Co-OFDMA}                   & \multicolumn{1}{l|}{-36.39}     & 0.64     & \multicolumn{1}{l|}{-15.07}          & 0.15        \\ \hline
\end{tabular}%
}
\end{table}


\section{Discussion and future work}
In this work we have shown a novel approach on multi-AP coordination in Wi-Fi networks, namely fine-grained Co-OFDMA using a fiber backhaul. It brings benefits especially for dense deployments in terms of latency and reliability, and when APs are not in each other's range, but serve STAs in overlapping areas. A design using the open-source openwifi and White Rabbit platforms is realized on software-defined radio. The performance of the implementation in terms of CFO pre-compensation accuracy and time synchronization is around $\pm$\SI{100}{Hz} and $\pm$\SI{48}{ns}, respectively.
The burst power of a two-user Co-OFDMA frame is increased by \SI{3}{dB} compared to individual two-user frames, while the EVM is improved by around \SI{1.4}{dB}. Hence, the implementation shows compliance to related physical-layer requirements and confirms the expected increase in received power and signal quality.

For future work, scheduling algorithms can be explored to efficiently allocate time and frequency resources using fine-grained Co-OFDMA.
Even if no dedicated over-the-air synchronization and coordination would be specified in the upcoming IEEE 802.11bn standard for (fine-grained) Co-OFDMA, we would recommend to consider including support for it in the preamble such that vendors can implement it using a dedicated backhaul. This work is an important step towards other MAPC schemes such as Co-BF and eventually JT, which we are investigating for future work using a similar methodology. 

\bibliographystyle{IEEEtran}

\bibliography{IEEEabrv,references.bib}

\end{document}